\documentclass[aps,prl,twocolumn,amsmath,amsfonts,floatfix,superscriptaddress]{revtex4-1}
\usepackage{natbib}
\usepackage{graphicx,amssymb}
\usepackage{color}

\newcommand{\revise}[1]{\textcolor{black}{{#1}}}
\newcommand{\rerevise}[1]{\textcolor{black}{{#1}}}

\begin{document}

\title{Ultrastable metallic glasses in silico}

\author{Anshul D. S. Parmar}

\affiliation{Laboratoire Charles Coulomb (L2C), Universit\'e de Montpellier, CNRS, 34095 Montpellier, France.}

\author{Misaki Ozawa}

\affiliation{Laboratoire de Physique Statistique, \'Ecole Normale Sup\'erieure, CNRS, PSL Research University, Sorbonne Universit\'e, 75005 Paris, France}

\author{Ludovic Berthier}

\affiliation{Laboratoire Charles Coulomb (L2C), Universit\'e de Montpellier, CNRS, 34095 Montpellier, France.}

\affiliation{Department of Chemistry, University of Cambridge, Lensfield Road, Cambridge CB2 1EW, United Kingdom.}

\begin{abstract}
We \revise{develop} a generic strategy and simple numerical models for multi-component metallic glasses for which the swap Monte Carlo algorithm can produce highly stable equilibrium configurations equivalent to experimental systems cooled more than $10^7$ times slower than in conventional simulations. This paves the way for a deeper understanding of thermodynamic, dynamic, and mechanical properties of metallic glasses. \revise{As first applications,} we \revise{considerably} extend configurational entropy measurements down to the experimental glass temperature, and demonstrate a qualitative \revise{change}  of the mechanical response of metallic glasses of increasing stability towards brittleness.
\end{abstract}

\maketitle

Glasses are obtained by cooling liquids  into amorphous solids~\cite{angell1995formation}. 
This process involves a rapidly growing relaxation time, making it difficult to investigate the nature of the glass transition in equilibrium~\cite{cavagna2009supercooled,berthier2011theoretical}.
%Glassy materials have many practical applications, such as food processing, optical fibers, and memory devices~\cite{berthier2016glass}.
Many types of materials can form glassy states, such as molecular, oxide, and colloidal glasses, having various practical applications~\cite{berthier2016glass}.
Among them, metallic glasses are a promising class known for higher strength and toughness~\cite{greer2013shear}, which is vital for applications.
Computer simulations represent a valuable tool to investigate glass properties with atomistic resolution~\cite{berthier2011theoretical}. Model metallic glasses are widely used because they are simpler than molecular liquids to understand the basic mechanisms of the glass transition \revise{and accompany practical applications}. However, typical cooling rates in silico are faster than in the laboratory by 6-8 orders of magnitude. Therefore, computer studies of metallic glasses may produce materials that behave differently from experimental systems. Our goal is to \revise{fill this wide gap} for metallic glasses in order to access thermodynamic, dynamic, and mechanical properties that can be directly compared to experiments.

\rerevise{Recently, the swap Monte Carlo algorithm has enabled the production of highly stable configurations for models of continuously polydisperse soft and hard spheres~\cite{berthier2016equilibrium,ninarello2017models}.} This was achieved by optimising the size distribution and pair interactions to produce good glass-formers (preventing crystallisation) with a massive thermalisation speedup~\cite{ninarello2017models}. It was however found that previous popular models for metallic glasses, such as the Kob-Andersen~\cite{kob1995testing} and Wahnstr\"om mixtures~\cite{wahnstrom1991molecular}, are either not well suited for the swap algorithm~\cite{flenner2006hybrid}, or crystallise too easily~\cite{brumer2004numerical,gutierrez2015static,ninarello2017models,ingebrigtsen2019crystallization,coslovich2018dynamic}. \rerevise{Further developments are clearly needed.} 

Here, \revise{we show how to develop} multi-component metallic glass-formers \revise{to benefit} from the dramatic speedup offered by swap Monte Carlo, and thus bridge the gap between \revise{metallic glass simulations} and experiments~\cite{yu2013ultrastable,yu2013ultrastable,aji2013ultrastrong,luo2018ultrastable,dziuba2020local}. Our strategy \revise{differs from earlier work~\cite{ninarello2017models} since it is} inspired by the microalloying technique used in metallic glass experiments~\cite{wang2007roles,gonzalez2016role}. We introduce additional species to the original binary Kob-Andersen mixture to simultaneously improve its glass-forming ability~\cite{tang2013anomalously,zhang2013computational} and swap efficiency~\cite{ninarello2017models}. \revise{This echoes the doping technique widely used in molecular liquids~\cite{angell1982test,takeda1999calorimetric,tatsumi2012thermodynamic} to prevent  crystallization~\cite{wang2007roles,gonzalez2016role,angell1982test,takeda1999calorimetric}.} The speedup provided by the swap Monte Carlo algorithm depends on the concentration of the doped species. For some models, we can produce \revise{for the first time} equilibrium configurations \revise{of metallic glasses} at the experimental glass transition temperature in silico. \revise{Our results pave the way for the next generation} of thermodynamic and mechanical studies of metallic glasses using computer simulations.

{\it Models---}The original Kob-Andersen (KA) model~\cite{kob1995testing} is a 80:20 binary mixture of $N_A$ Lennard-Jones particles of type A, and $N_B$ particles of type B, mimicking the mixture Ni-P. We add a new family of particles, of type C, which can be a single type (ternary mixture) or several types (multi-component). The pair interaction is 
\begin{equation}
v_{\alpha_i \beta_j}(r) = 4\epsilon_{\alpha_i \beta_j} \left[ \left( \frac{\sigma_{\alpha_i \beta_j} }{r}  \right)^{12} - \left( \frac{ \sigma_{\alpha_i \beta_j} }{r} \right)^{6} \right],
\end{equation}
where $\epsilon$ and $\sigma$ are the energy scale and interaction range, respectively. We specify the particles index by Roman indices and the family type by Greek indices. The potential is truncated and shifted at the cutoff distance $r_{{\rm cut},ij} = 2.5 \sigma_{\alpha_i \beta_j}$. For particles A and B, we use the interaction parameters of the original KA model:
$\epsilon_{AB}/\epsilon_{AA}=1.5$, $\epsilon_{BB}/\epsilon_{AA}=0.5$, 
and $\sigma_{AB}/\sigma_{AA}=0.8$, $\sigma_{BB}/\sigma_{AA}=0.88$. Energy  
and length are in units of $\epsilon_{AA}$ and $\sigma_{AA}$, 
respectively. \revise{Given the large size and energy disparities, performing particles swaps between A and B particles is prohibited~\cite{flenner2006hybrid} which leaves the standard KA model out of the recent swap developments.}

We introduce $N_{C}$ particles of type C. Each C particle is characterized by a continuous variable $\omega_i \in [0,1]$ so that its interactions with A and B particles are given by 
\begin{eqnarray}
	 X_{AC} &=& \omega_i X_{AA} + (1-\omega_i) X_{AB}, \nonumber \\ 
	X_{BC} &=& \omega_i X_{AB} + (1-\omega_i) X_{BB}, 
	\label{eq:def_X1}
\end{eqnarray}
where $X$ stands for both $\epsilon$ and $\sigma$, so that C particles are identical to A (B) particles when $\omega_i=1$ (0) and smoothly interpolate between both species for $0< \omega_i< 1$. Two C particles $i$ and $j$ interact between each other additively: 
\begin{equation}
 	X_{C_i C_j}  =  \omega_{i j} X_{AA} + (1-\omega_{i j}) X_{BB},
 	\label{eq:def_X2}
\end{equation}
where $\omega_{ij}=(\omega_i + \omega_j)/2$.

\begin{figure}
\includegraphics[width=8.5cm]{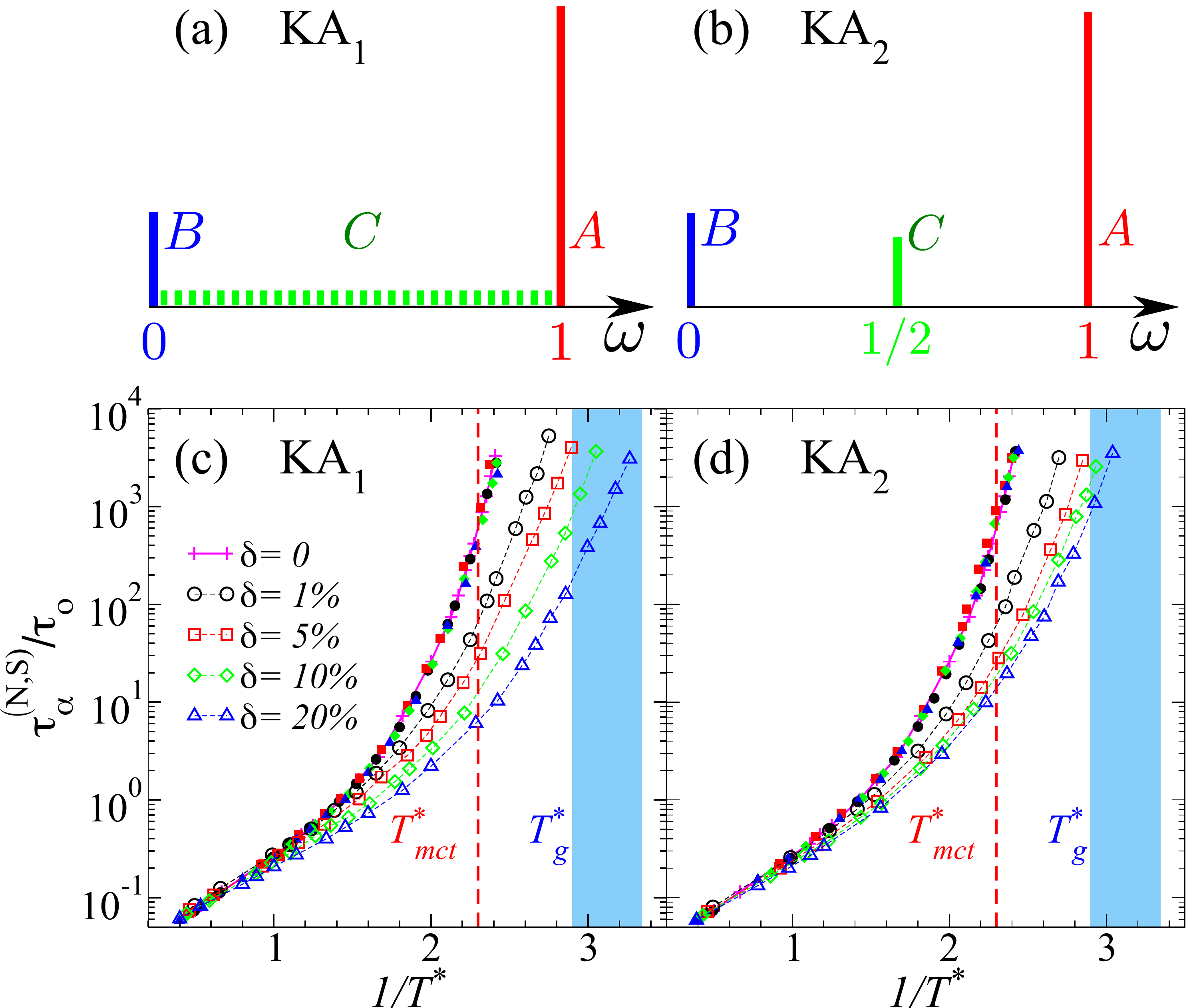}
\caption{(a,b) Two families of Lennard-Jones models, composed of A and B particles interacting as in the KA model, and C particles intermediate between A and B types depending on the variable $\omega$. 
%KA$_1$ (a) and KA$_2$ (b) are obtained by a flat $P(\omega)$ and $P(\omega)=\delta(\omega-1/2)$, respectively. 
The C particles improve the glass-forming ability and the efficiency of the swap Monte Carlo algorithm. Relaxation times for the (c) KA$_1$ and (d) KA$_{2}$ models using full ($\tau_{\alpha}^{(N)}$) and empty ($\tau_{\alpha}^{(S)}$) symbols. The blue box indicates the extrapolated location of the experimental glass transition temperature $T_{g}^*$, and
$T_{mct}^*$ is the mode-coupling crossover.}
\label{fig1}
\end{figure}

This generic framework offers multiple choices for the distribution of C particles, depending on the parameters $N_C$ and on the chosen distribution $P(\omega)$ of the variable $\omega$. We have explored two simple families, illustrated in Figs.~\ref{fig1}(a,b). The first family, KA$_1$, is obtained using a flat distribution $P(\omega)$ on the interval $[0,1]$, see Fig.~\ref{fig1}(a). This corresponds to a multi-component system where C particles continuously interpolate between A and B components. The second family, KA$_2$, is obtained by taking the opposite extreme where $P(\omega)=\delta(\omega-1/2)$, see Fig.~\ref{fig1}(b). In that case, we simulate a discrete ternary mixture. In both cases, we define $\delta = N_C/(N_A+N_B)$ and consider a range of $\delta$ values from $\delta=0\%$ (original KA mixture), up to $\delta=20\%$. \revise{Contrary to previous work~\cite{ninarello2017models}, the size dispersity quantified by the variance of the diameter distribution is nearly constant across KA, KA$_1$ and KA$_2$ models. We perform simulations in a periodic cubic cell of volume $V$ in three dimensions. All models are simulated at the number density $\rho=1.2$, denoting the number of particles in unit volume $\sigma_{AA}^{3}$}.

{\it Swap Monte Carlo algorithm--}To achieve equilibration at very low temperatures, we perform Monte Carlo (MC) simulations possessing both translational displacements and particle swaps~\cite{frenkel2001understanding,grigera2001fast}. For the normal MC moves, a particle is randomly chosen and displaced by a vector randomly drawn within a cube of linear size $\delta r_{\rm max} = 0.15$. The move is accepted according to the Metropolis acceptance rule, enforcing detailed balance. Such MC simulations show quantitative agreement with molecular dynamics simulations in terms of glassy slow dynamics~\cite{berthier2007monte}. 

When using swap MC, we also perform particle swaps. We randomly choose a C particle, say particle $i$, characterised by $\omega_i$. We then randomly choose a value $\Delta \omega$ in the interval $\Delta \omega = \pm 0.8$ and choose a second particle within this interval, say particle $j$. We estimate the energy cost to exchange the type of the two particles, $\omega_i \leftrightarrow \omega_j$, and accept the swap according to the Metropolis rule. In the swap MC scheme, we perform swap moves with probability $p=0.2$, and translational moves with probability $1-p=0.8$. All parameters, $(\delta r_{max}, p, \Delta \omega)$ have been carefully optimised to maximise the swap efficiency, see supplementary material (SM~\footnote{See Supplemental Material at [url] for details about simulation methods, temperature scaling, determination of the experimental glass transition temperature, thermodynamics and its relation with dynamics, glass-forming ability, and an additional model system, which includes Refs.~\cite{sastry2000liquid,sastry1997statistical,ferry1980viscoelastic,ozawa2019does,allen2017computer,adam1965temperature,kirkpatrick1989scaling,lubchenko2007theory,honeycutt1987molecular}
}). In particular, swaps with larger $\Delta \omega$ are essentially all rejected, confirming that direct A $\leftrightarrow$ B swaps are impossible.
In essence, the C particles thus allow two-step exchanges, such as A $\leftrightarrow$ C $\leftrightarrow$ B.
%This is the core mechanism of our strategy, which is achieved by specific tuning of the doped species given by Eqs.~(\ref{eq:def_X1}) and (\ref{eq:def_X2}) so that the energy cost of swaps involved by the C particles is reduced.
Although we only apply this strategy to the KA model, we expect that it should generically apply to high-entropy alloys which have more than five components~\cite{zhang2014microstructures}. In both normal and swap MC schemes, one Monte Carlo time step represents $N$ attempts to make an elementary move. Timescales are reported in this unit.

{\it Glass-forming ability--}Thanks to modern computer resources, the original KA model is now found to be prone to crystallisation~\cite{toxvaerd2009stability,coslovich2018dynamic,ingebrigtsen2019crystallization}. We have repeated the detailed common neighbor analysis of Ref.~\cite{coslovich2018dynamic}. \revise{We detected no sign of crystalline environments in our extended models, KA$_1$ and KA$_2$, across the wide temperature regime where thermalisation can be achieved using the swap MC algorithm, see SM. Thus, the extended KA models developed here are much better glass-formers than the original KA model.
%We find that the inclusion of the C particles enhances the glass-forming ability of the models, since no sign of crystallisation is detected. 
Similarly to experiments, the doping C particles considerably frustrate the system against crystallisation~\cite{wang2007roles,gonzalez2016role,angell1982test,takeda1999calorimetric}.}
 
{\it Equilibration speed-up--}The relaxation time $\tau_\alpha$ of the system is quantified from the time decay of the self-intermediate scattering function for all particles, $F_s(q,t=\tau_\alpha)=1/e$. We use $q=7.34$, close to the first diffraction peak of the static structure factor. We respectively denote $\tau_{\alpha}^{(N)}$ and $\tau_{\alpha}^{(S)}$ the relaxation times for the normal (N) and swap (S) dynamics.
We finally rescale the relaxation times using its value $\tau_o = \tau^{(N)}_\alpha(T=T_o)$ at the onset temperature $T_o$ at which the relaxation time starts to deviate from the Arrhenius law. %\MO{$\tau_o$ is normal or each model?? There is no explanation in both main and SM.)}

We first concentrate on the physical dynamics using normal MC simulations for both models, KA$_1$ and KA$_2$, and various values of $\delta$. We find that the temperature dependence of $\tau_\alpha^{(N)}$ for all models is very similar, and is only weakly affected by the C particles \revise{(see SM)}. 
%Since we work at constant number density $\rho$, the C particles necessarily affect the pressure, which may thus perturb the dynamics. 
The presence of the C particles changes the energy or temperature scale from the original model. To account for this perturbation \revise{and ease the comparison between models}, we introduce a rescaled temperature, $T^*=T(1 + \varepsilon(\delta))$, such that the data $\tau_\alpha^{(N)}$ versus $1/T^*$ for all models coincide, see Fig.~\ref{fig1}(c,d) \revise{(see also the $\tau_\alpha^{(N)}$ versus $1/T$ plots in SM)}. 
The measured $\varepsilon$ values reported in Table~\ref{data} are small and compatible with a linear growth, $\varepsilon(\delta) \simeq \delta$, suggesting that C particles simply act as a linear thermodynamic perturbation (see SM for KA$_{2}$). We confirm in SM that the pair structure is also weakly affected. 
%The radial distributions for different $\delta$ are superimposed at $T^*$. 
From now on, we use the temperature scale $T^*$ and thus, by definition, all models \revise{studied in this paper} display the same physical \revise{(normal MC)} dynamics \revise{as a function of $T^*$. They have the same reference temperatures as the original KA model:} their onset temperature is $T_o^* \simeq 0.7$, and the mode-coupling crossover is at $T_{mct}^* \simeq 0.435$. These conventional MC simulations can access $\tau_\alpha^{(N)} /\tau_o \sim 10^4$ for $N=10^3$ particles, corresponding to the lowest simulated temperaure $T_{low}^* = 0.415$ and about 10 days of CPU time. Following earlier work~\cite{ninarello2017models}, we locate the experimental glass transition temperature $T_g^*$ by extrapolating the measured dynamical date towards $\tau_\alpha^{(N)}/\tau_0=10^{12}$ using various functional forms which provide a finite range for its location. We find $T^*_g \in [0.3 - 0.345]$, see  Fig.~\ref{fig1}, and suggest $T_g^*\approx 0.3$ as our favored estimate obtained using the parabolic law~\cite{elmatad2010corresponding}.

\begin{table}
\begin{tabular}{l*{5}{c}}
KA$_{1}$              & \quad $\delta=0\%$  \quad & \quad 1\% \quad & \quad 5\% \quad & \quad 10\% \quad & \quad 20\% \\
\hline
$\varepsilon$ 					 				& $0$ & $0.01$ & $0.08$ & $0.13$ & $0.25$ \\
$T^{*}_{low}/T^{*}_{mct}$		        & $0.954$ & $0.824$ & $0.787$ & $0.753$ & $0.704$ \\
Speedup & $1$ & $10^{2}$ & $6\times 10^{3}$ & $8\times 10^{4}$ & $2\times 10^{7}$ \\
$K_{T}$            		 					& $0.335$ & $0.334$ & $0.312$ & $0.293$ & $0.274$ \\
$T^{*}_{K}$					 				& $0.252$ & $0.246$ & $0.236$ & $0.224$ & $0.210$ \\
\hline
\end{tabular}
\caption{Characteristics of the KA$_{1}$ models. Scaling factor for the temperature $\varepsilon$, the lowest simulated temperature $T^*_{low}$ relative to the mode-coupling crossover $T^*_{mct}$, thermalisation speedup, thermodynamic fragility $K_T$, and extrapolated Kauzmann temperatures $T^{*}_K$.} 
\label{data}
\end{table} 

Our \revise{first important achievement} follows from the temperature evolution of the relaxation times when using swap MC in Fig.~\ref{fig1}. Whereas the original KA model with $\delta=0$ can be thermalised down to $T_{low}^* \approx 0.415$, we find that thermalisation is achieved at much lower temperatures as soon as $\delta>0$, with a speedup that {\it increases continuously and exponentially fast} with $\delta$. For an equivalent numerical effort, we find for $\delta=1\%-20\%$, $T^*_{low} =0.306-0.358$ (for KA$_1$) and $T^*_{low} =0.326-0.371$ (for KA$_2$). The lowest temperature corresponds to $T^*_{low} \approx 0.7 T_{mct}^* \approx T_g^*$. Converting these temperatures into timescales, we estimate that the numerical speedup varies from a factor $10^2$ for $\delta=1\%$, up to more than $10^7$ for $\delta=20\%$. Thus, even a small amount of doping has a massive impact on the swap efficiency. \revise{The proposed metallic glass models considerably widen the accessible temperature regime available to computer simulations, without suffering from crystallizations.}

\begin{figure}
\includegraphics[width=8.5cm]{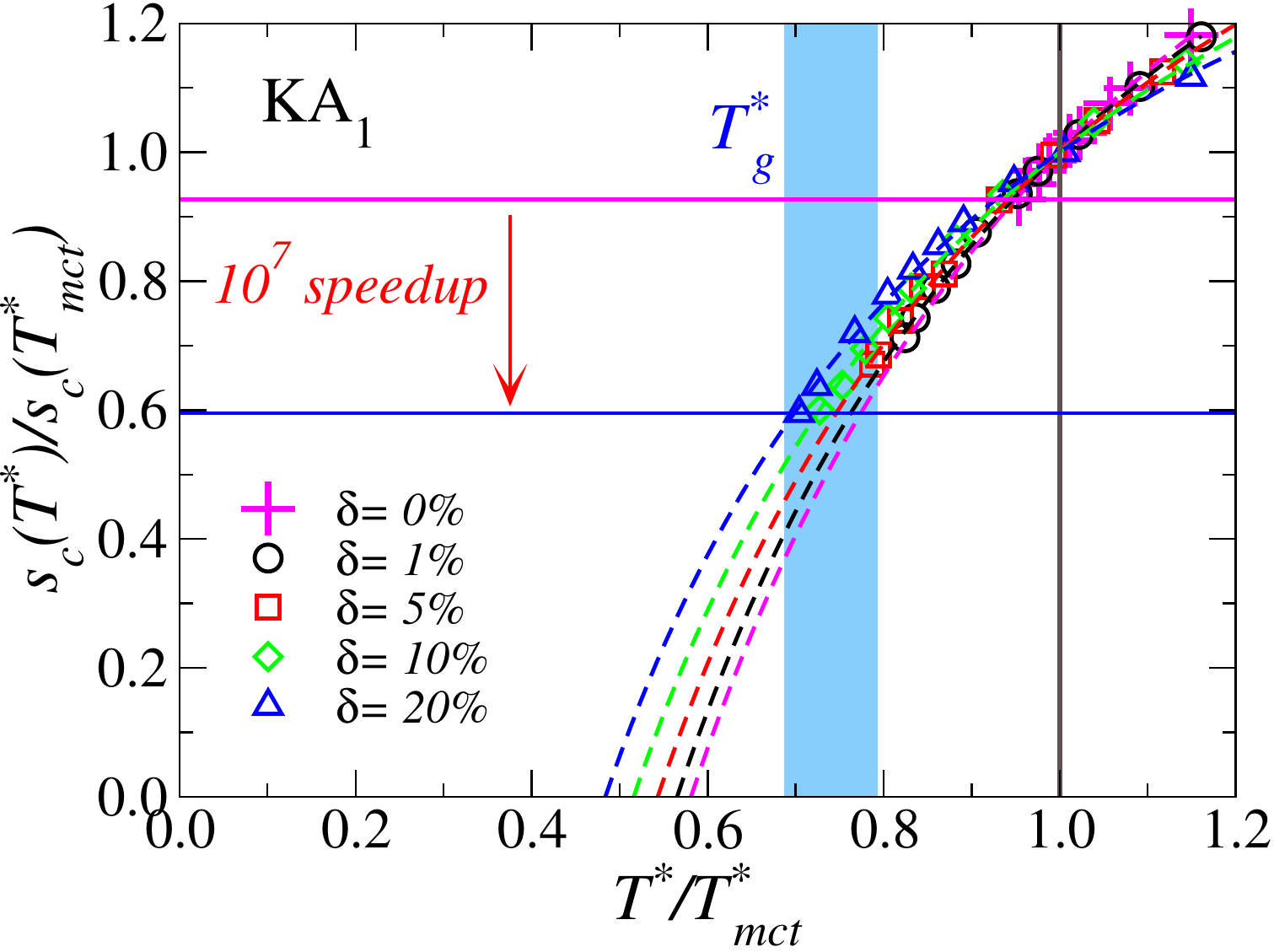}
\caption{Temperature evolution of configurational entropy using variables rescaled at the mode-coupling crossover temperature $T^{*}_{mct}$.
The horizontal lines correspond to the lowest accessible temperature for the original KA model and the $\delta=20\%$ KA$_1$ model.
}
\label{fig:sc}
\end{figure} 

{\it Configurational entropy--}We now characterise the configurational entropy, $s_c(T^*)$, of the very low temperature states produced with swap MC. We determine the configurational entropy from its conventional definition, $s_{c}(T^{*}) = s_{tot}(T^{*}) - s_{vib}(T^{*})$ \cite{kauzmann1948nature,sciortino1999inherent,sastry2000evaluation,berthier2019configurational}. The equilibrium entropy, $s_{tot}$, is straightforwardly measured by thermodynamic integration from the ideal gas to the studied state point~\cite{sastry2000evaluation}. The
vibrational entropy, $s_{vib}$, is obtained by a constrained Frenkel-Ladd~\cite{frenkel1984new} thermodynamic integration, generalised to properly quantify the mixing entropy contribution to the vibrational entropy~\cite{ozawa2018configurational}. This is a crucial point for the present models where polydispersity changes continuously with $\delta$, and
alternate approaches, for instance, using inherent structures, would be inadequate~\cite{ozawa2017does}. Figure~\ref{fig:sc} shows the temperature evolution of the configurational entropy of KA$_{1}$ models. We use $T_{mct}^*$ as a useful temperature scale to normalise both temperatures, $T^*/T_{mct}^*$, and entropies, $s_c(T^*)/s_c(T^*_{mct})$, in the spirit of Kauzmann~\cite{kauzmann1948nature}. Our data for the KA model are consistent with previous work~\cite{banerjee2016effect}, and stop at $T_{low}^* / T^*_{mct} \approx 0.954$ where $s_c(T^{*}_{low})/s_{c}(T^{*}_{mct}) \approx 0.93$, corresponding to the deepest states accessible with a conventional MC algorithm. Figure~\ref{fig:sc} shows that the thermalisation speedup obtained by increasing $\delta$ is accompanied by a strong reduction of the configurational entropy. Hence, the deeply supercooled states obtained using swap MC correspond to state points where much fewer amorphous packings are available to the system, which should translate into a larger point-to-set correlation length~\cite{bouchaud2004adam,berthier2017configurational,berthier2019zero}. In earlier studies of the KA model, the putative Kauzmann transition was determined by fitting the decrease of $s_c$ with the empirical form $s_c = A (1 - T^*_K/T^*)$, which also allows the determination of the thermodynamic fragility: $K_T \equiv A T^{*}_K$~\cite{sastry2001relationship}. We extend this analysis to KA$_1$ models and report $T_K$ and $K_T$ in Table~\ref{data}. Both quantities show a modest, but systematic decrease with $\delta$. Remarkably, \revise{increasing the studied time window from 4 to 11 orders of magnitude (from $\delta=0\%$ to 20\%), the steep temperature dependence of $s_c$ remains consistent} with an entropy crisis taking place at $T_K \approx 0.5 T^*_{mct}>0$. \revise{In particular, we detect no sign of a new mechanism to `avoid' it~\cite{royall2018race}.} 
These data also contradict the \revise{arguments} that models where swap MC works well are qualitatively distinct from those where it does not~\cite{wyart2017does,berthier2019can,lubchenko2017aging}, \revise{and constitute our second important result.}

\begin{figure}
\includegraphics[width=8.5cm]{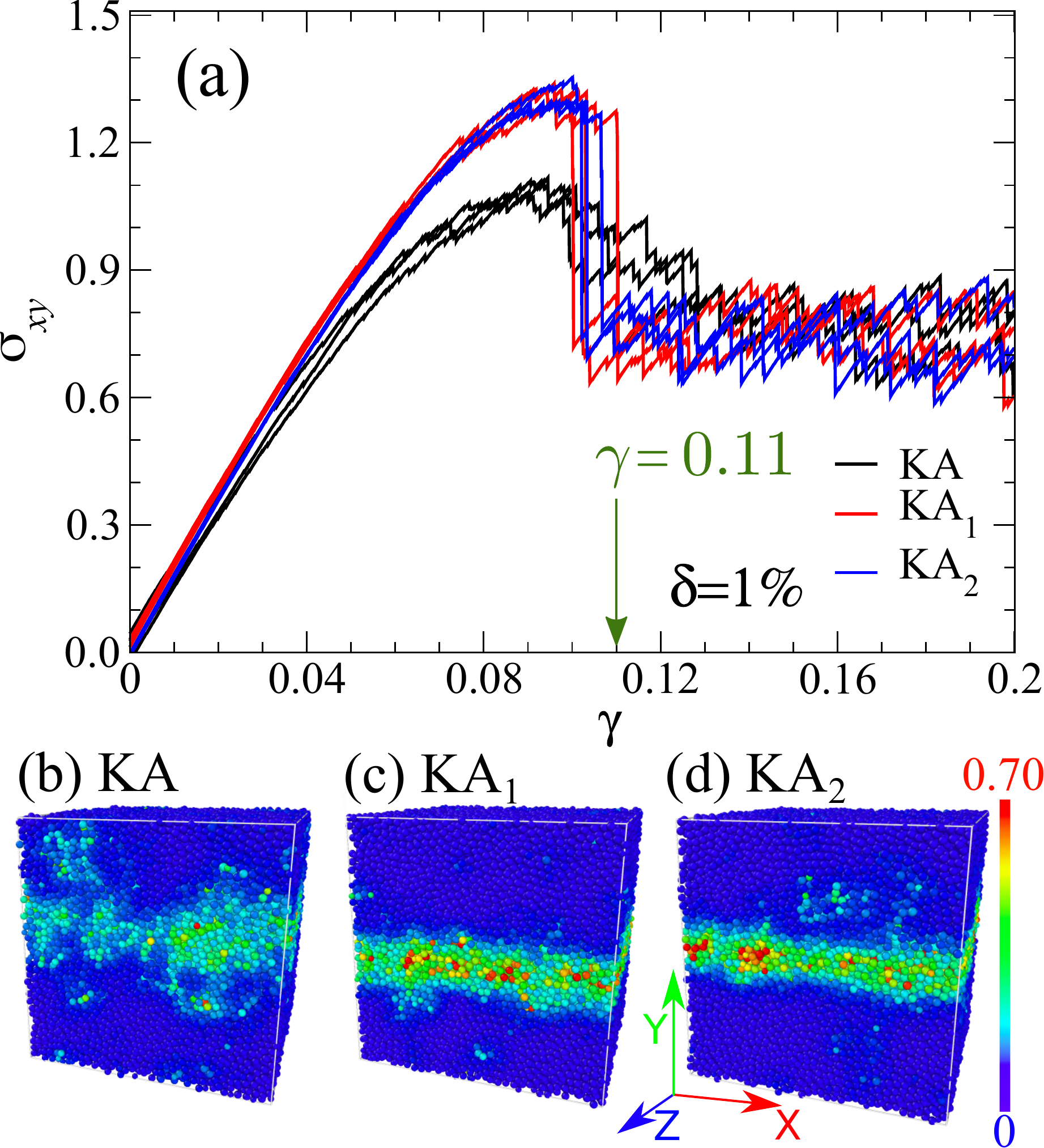}
\caption{(a) Stress-strain curves for three models, KA, KA$_1$, and KA$_2$ with $\delta=1 \%$. For each model, three individual realizations with $N=50000$ are shown. (b,c,d) Snapshots of non-affine displacement between $\gamma=0$ and $\gamma=0.11$ (vertical arrow in (a)) for KA (b), KA$_1$ (c), and KA$_2$ (d).} 
\label{shear}
\end{figure}

{\it Brittle yielding--}Turning to rheology, we demonstrate that accessing highly stable glassy configurations qualitatively affects how simulated metallic glasses yield. \revise{It was recently suggested that glass stability induces a ductile-to-brittle transition, confirmed numerically in a model for soft glasses~\cite{ozawa2018random,ozawa2020role,yeh2020glass}. Here, we establish that a similar transition exists also in metallic glasses.} To this end, we consider a larger system size, $N=5 \times 10^4$, and apply the following preparation protocol for the original KA model, and the $\delta=1\%$ KA$_1$ and KA$_2$ models. First, we thermalise the system at high temperature, $T^*=2.0$. 
%Second, we instantaneously quench the temperature  to state points where $\tau_\alpha^{(S)} \simeq 10^{10}$~MC steps ($T^*=0.373$ and $0.319$ for KA and KA$_{1,2}$, respectively), and let them age during $10^6$ swap MC steps. 
Second, we instantaneously quench to the temperature $T^*=0.373$ and $0.319$ for KA and KA$_{1,2}$, respectively, where $\tau_\alpha^{(S)} \simeq 10^{10}$~MC steps, and let them age during $10^6$ swap MC steps. 
We expect to produce an ordinary computer glass of modest stability for the KA model, but very stable configurations for KA$_1$ and KA$_2$ models. These aged glasses are quenched to $T=0$, and sheared using a strain-controlled athermal quasi-static protocol~\cite{maloney2006amorphous}. We apply a uniform shear along the xy-plane, with strain increments $\Delta \gamma = 10^{-4}$. We measure the xy-component of the shear stress, $\sigma_{xy}$, to obtain the stress-strain curves shown in Fig.~\ref{shear}(a). We visualise non-affine particle displacements~\cite{falk1998dynamics} in Fig.~\ref{shear}(b-d).   
 
For the KA model, the stress shows an initial quasi-linear increase with small plastic events, a stress overshoot punctuated by many larger plastic events, and a gradual approach to steady-state. Near yielding, plasticity is spatially heterogeneous but spreads over the entire system, see Fig.~\ref{shear}(b), in agreement with previous findings~\cite{utz2000atomistic,fan2017effects,rodney2011modeling}. For stable initial configurations, the stress overshoot transforms into a unique, sharp, macroscopic stress discontinuity, see Fig. \ref{shear}(a). This brittle behavior is accompanied by a clear system-spanning shear-band, see Fig.~\ref{shear}(c,d). The tendency to shear localisation upon increasing stability is well-documented~\cite{varnik2003shear,shi2005strain}, but a genuine non-equilibrium discontinuous yielding transition only occurs for highly stable glassy systems~\cite{ozawa2018random}, see also Refs.~\cite{ketkaew2018mechanical,kapteijns2019fast,bhaumik2019role}. \revise{These results extend to an experimentally relevant class of materials the observation that brittle yielding and macroscopic shear-band formation can be studied in atomistic simulations. They also suggest the universality of the random critical point controlling the brittle to ductile transition~\cite{ozawa2018random}. This constitutes our third important result.}

{\it Perspectives--}The multi-component models for metallic glasses developed here can be efficiently thermalised via swap Monte Carlo simulations down to temperatures that are not currently accessible to conventional simulation techniques \revise{and are comparable to the experimental glass transition.}
These models fill the gap between experimental and numerical works. Considering the extensive use made of the KA model~\cite{kob1995testing}, \rerevise{the improved glass-forming ability and thermalisation efficiency will stimulate many future studies.} Immediate applications concern further analysis of thermodynamic, dynamical, and mechanical properties of the stable configurations obtained here, to address questions regarding the Kauzmann temperature, the validity of the Adam-Gibbs relation (see SM for an initial attempt) and a microscopic description of shear band formation and failure in metallic glasses. More broadly, the strategy proposed here is simple and versatile, and can certainly be improved further. For example, increasing the number of components and their concentration allows the system to reach below $T_g$ very efficiently (see SM for a four-component model). It could also be used to model some specific multi-component materials \revise{(for instance of the Ni-Pd-P type)} and high-entropy alloys, to deepen our theoretical understanding of metallic glasses and help the design of novel materials with specific properties.

\acknowledgments
We thank D. Coslovich and A. Ninarello for discussions and sharing data and codes. We also thank C. Cammarota, A. Liu and S. Sastry for useful discussions. This work was supported by a grant from the Simons Foundation (\#454933, L. Berthier).

\bibliography{metallic-glasses.bib}{}

\newpage

\section{Supplementary Material}

In the supplementary material, we provide additional information regarding the following aspects: (i) details of the simulation methods, (ii) comparison of normal dynamics and static correlation function for various $\delta$ values, (iii) estimating the experimental glass transition temperature, (iv) details of the computation of entropies and KA$_2$ model, (v) test of the Adam-Gibbs relation, (vi) robustness against crystallization.

\section{Simulation methods}

Translational Monte Carlo moves are conceptually straightforward and require only optimisation of the time decay of the structural relaxation time~\cite{frenkel2001understanding}. We find that picking a random displacement within a cube of linear length $\delta r_{max}$ $=0.15$ with the Metropolis acceptance criteria makes the dynamics optimal. These dynamics can be be viewed as equivalent to a discretised Brownian dynamics. One Monte-Carlo sweep consists of $N$ such elementary trial moves, and timescales are reported in this unit~\cite{berthier2007monte}.

\begin{figure}[htp]
\centering{}
\includegraphics[width=8.5cm]{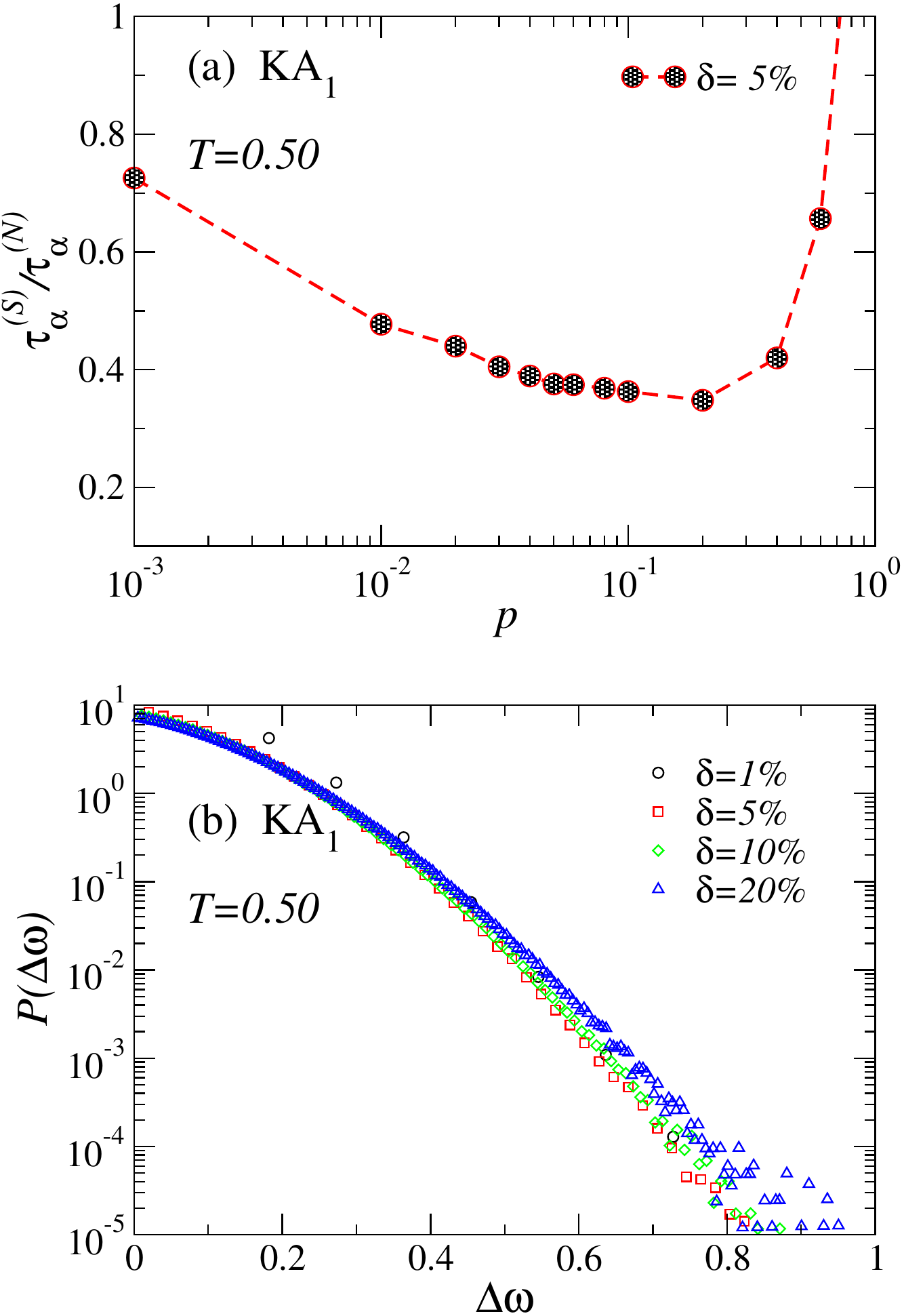}
\caption{(a) The relaxation time $\tau_{\alpha}^{(S)}$ of the $KA_1$ model as a function of the probability of swap moves $p$. (b) The swap acceptance rate, $P(\Delta \omega)$, as a function of particle disparity $\Delta \omega=|\omega_{i}-\omega_j|$ between the particle pairs for which the swap is attempted.}
\label{fig:optimal_swap}
\end{figure}

In addition to the conventional Monte-Carlo moves, the swap Monte Carlo approach consists of adding particle swaps, where the value of $\omega$ of a randomly drawn pair of particles is exchanged. We mix translational and swap move with probability $p$ and $1-p$, respectively. As shown in Fig.~\ref{fig:optimal_swap}(a), the optimal choice is near $p\approx 0.2$, for which structural relaxation decays the fastest~\cite{ninarello2017models}.

To further optimize the swap Monte-Carlo moves, we measure the acceptance probability $P(\Delta \omega)$ as a function of the disparity of particle pairs quantified by $\Delta \omega = | \omega_i - \omega_j |$. We consider the temperature $T=0.50$ and $\delta$ from 1\% to 20\%. In Fig.~\ref{fig:optimal_swap}(b), we see that acceptance is larger when $\Delta \omega$ is small and particles thus strongly resemble each other. The acceptance fall to $10^{-5}$ when $\Delta \omega \approx 0.8$, showing that direct exchanges between A and B particles are strongly suppressed. The trade-off is that larger $\Delta \omega$ more efficiently thermalise the system, but are less frequently accepted. We fix a threshold $\Delta \omega = 0.8$ and do not attempt swapping particles for larger values.
For a given C particle, we pick a direction for $\pm 0.80$ with equal chance, and randomly pick another particle can lie beyond $\omega$ = 1(A) or $\omega$ = 0(B) range as well, in that case, we randomly pick particle of type A or B

Dynamics is characterized by the alpha-relaxation times obtained by the self-intermediate scattering function. Since the particle's type changes throughout the swap simulations, the relaxation time is calculated for the complete system, as the time at which the self-intermediate scattering function $F_s({\bf q}, t)$, for the first peak of static structure factor, decays to a value of $1/e$. $F_{s}(q, t)$ is calculated using
\begin{equation}
F_{s}(q, t) = \frac{1}{N}\left < \sum_{i=1}^{N} \exp \left[ -i {\bf q} \cdot ({\bf r}_i (t) - {\bf r}_i(0)) \right] \right > ,
\end{equation}
where ${\bf r}_{i}(t)$ are the positions of particle $i$ at time $t$ and the
averaging is performed over many times origins.

\section{Temperature scaling for various models}

\begin{figure}[]
\centering{}
\includegraphics[width=8.5cm]{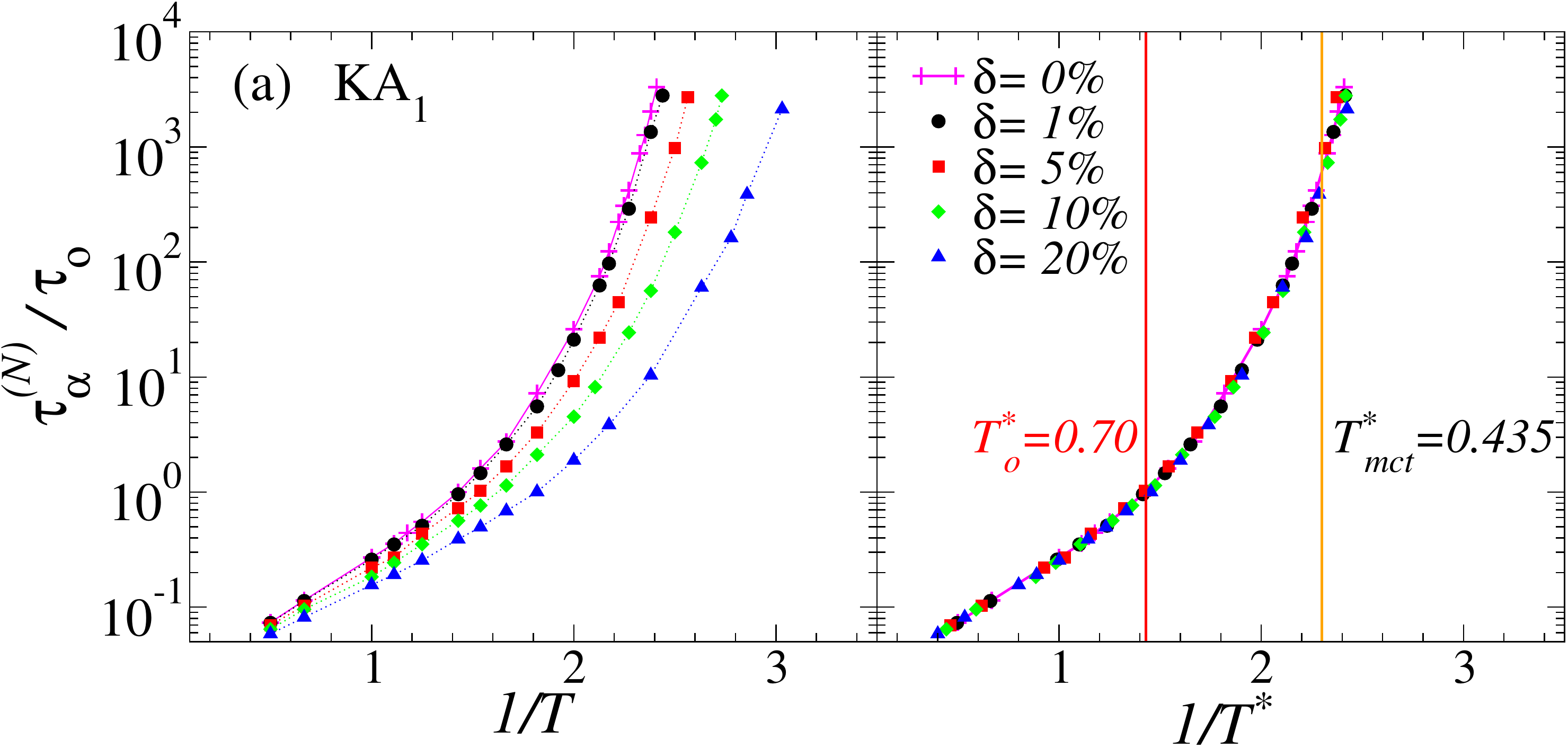}
\includegraphics[width=8.5cm]{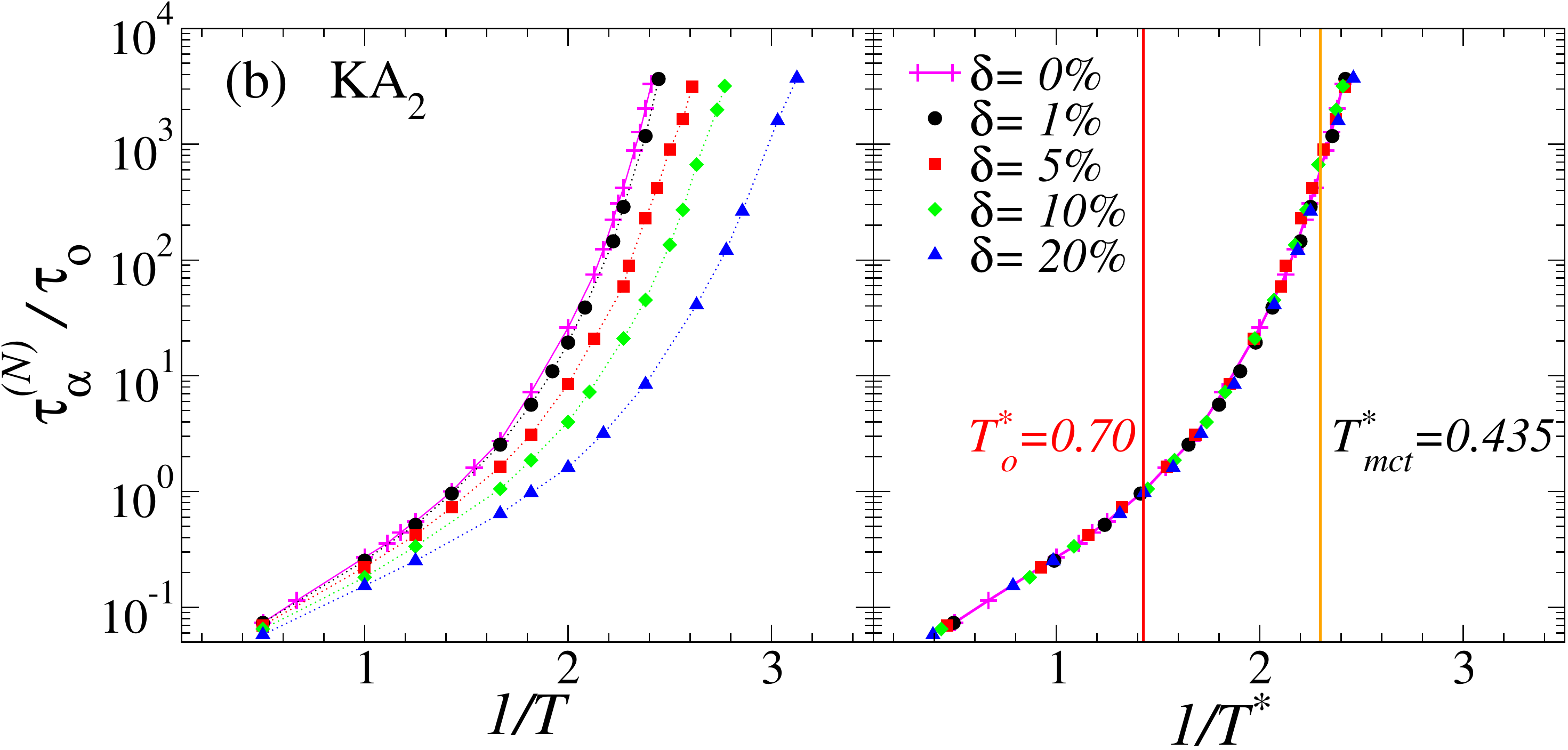}
\caption{Relaxation times for the KA$_1$ model (a) and KA$_2$ model (b) using either the temperature, $T$ (left panels) or rescaled temperature $T^*$ (right panels).} 
\label{scaling}
\end{figure}

The introduction of additional C particles represents a small thermodynamic perturbation to the original model, even if we fix the number density to $\rho=1.2$ for all $\delta$ values. This can be seen in Fig.~\ref{scaling} where we show the evolution of $\tau_\alpha^{(N)}$ for the normal dynamics as a function of $1/T$. A systematic shift of the data with $\delta$ is clearly visible.

However, by scaling the temperature with a single constant, $T^{*} = T(1+\epsilon(\delta))$, the dynamics for all models can be rescaled on a single master curve, as shown in Fig. \ref{scaling}. The scaling works for both KA$_1$ and KA$_2$ models. The temperature $T^*$ is thus useful to compare different systems between each other. By definition $T^*=T$ for the original KA model with $\delta=0\%$, so that the relevant temperature scales for the KA model directly apply to KA$_1$ and KA$_2$ models.

\begin{figure}[tp]
\centering{}
\includegraphics[scale=0.5]{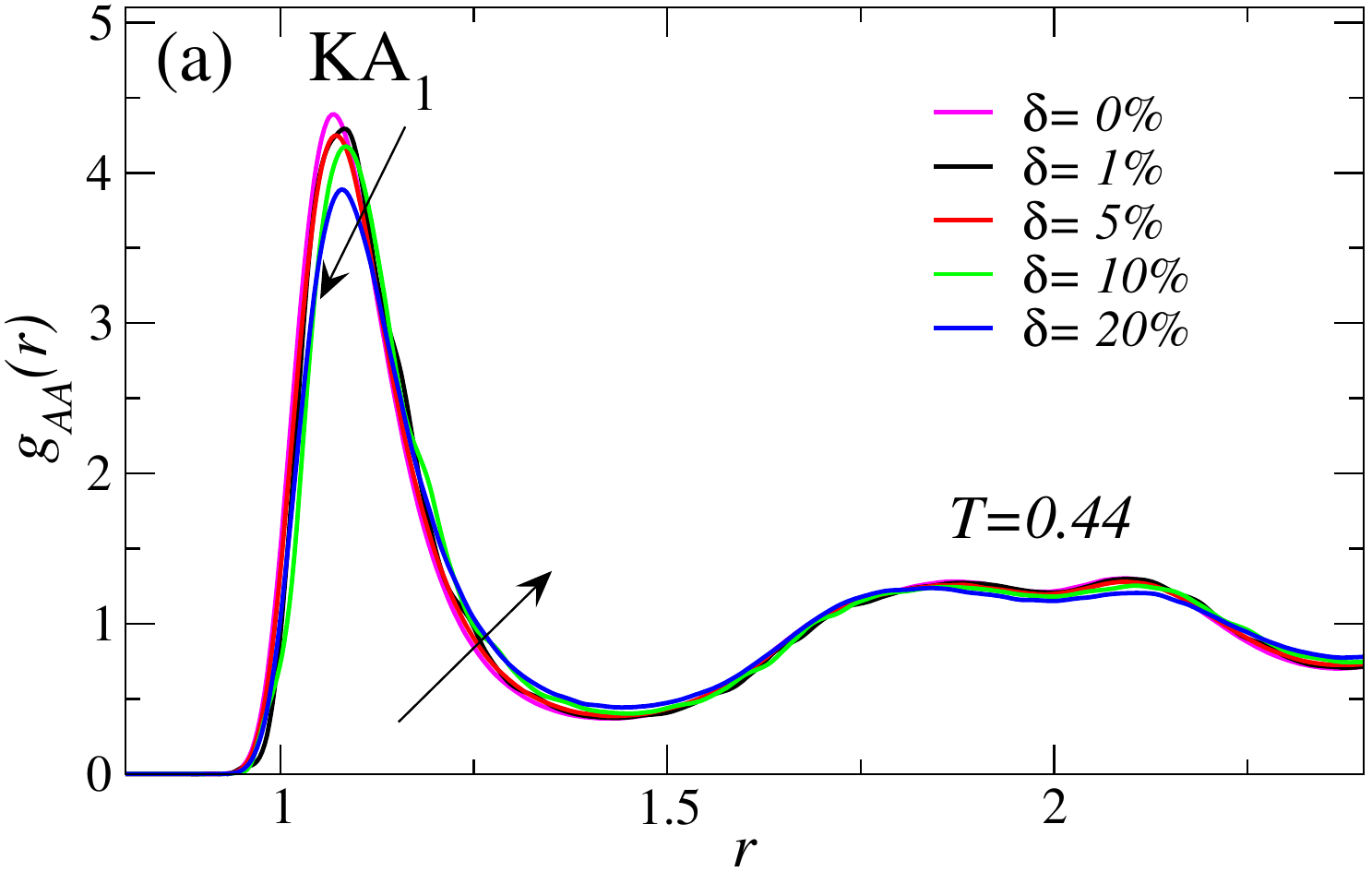}
\includegraphics[scale=0.5]{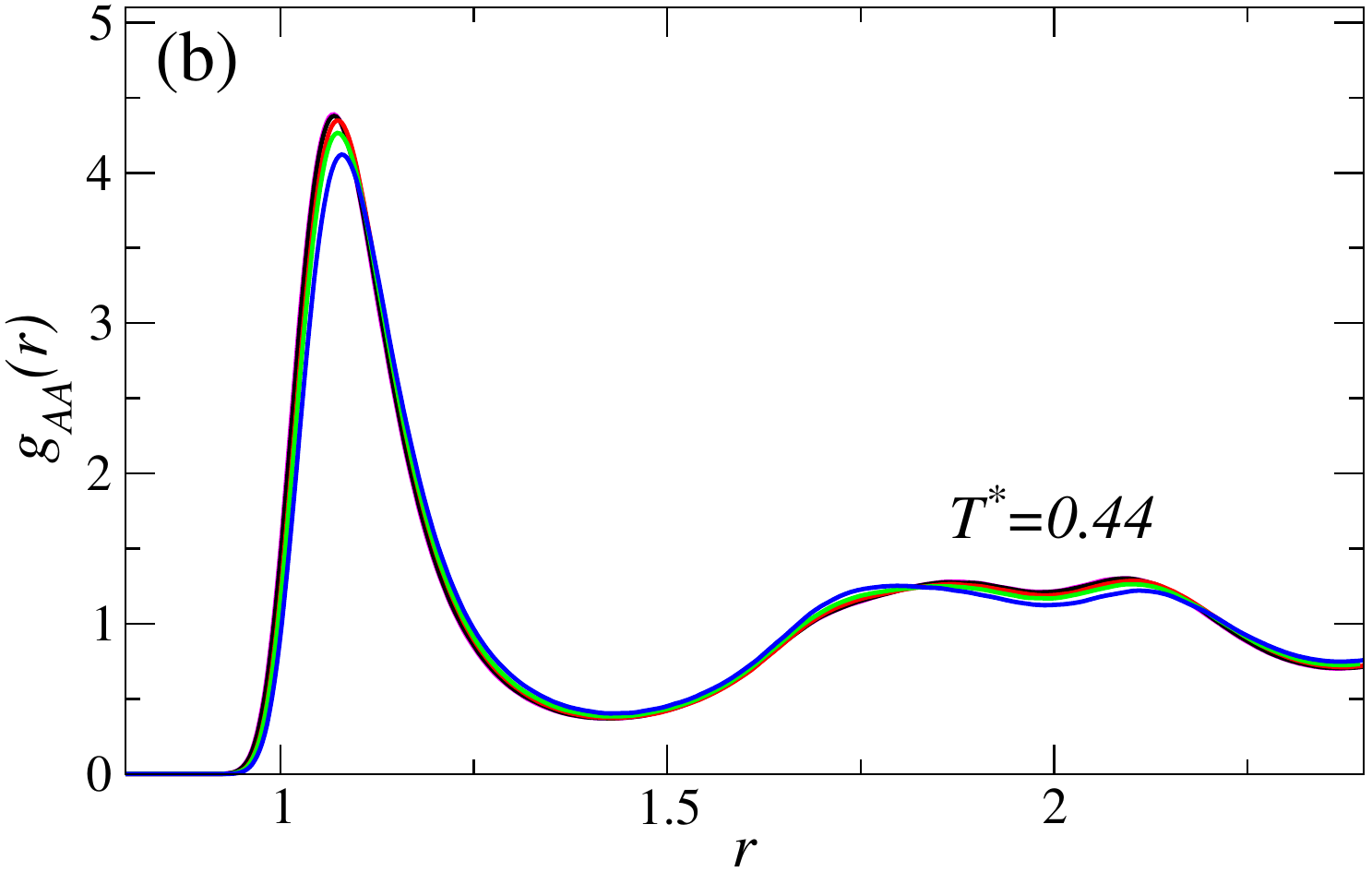}
\caption{Pair correlation function for the majority specie A for the KA$_1$ models at (a) $T=0.44$, (b) $T^*=0.44$.} 
\label{g(r)}
\end{figure}

The temperature shift quantified by $\epsilon(\delta)$ is largely due to the thermodynamic perturbation induced by the doping C particles. To show this, we present in Fig.~\ref{g(r)}(a) the evolution of the radial distribution function for the majority species, $g_{AA}(r)$ at a given temperature, $T=0.44$ and increasing $\delta$ values. Clearly, the pair correlation function is weakly affected by the addition of C particles, in a way that increases with $\delta$. Rescaling the temperature to make the dynamics coincide removes a large part of this change, but not all of it. This is shown in Fig.~\ref{g(r)}(b) where the data for $g_{AA}(r)$ are now collected for a given value of $T^*=0.44$. The fact that a small change of the pair correlation survives the temperature rescaling shows that despite the models have a very similar dynamics, they can still be distinguished even at the level of the pair structure. They are, therefore, not strictly identical.
The improved glass-forming ability (see below) would stem from this slight difference of structure.

It has been reported that the standard KA mixture shows a gas-liquid (or gas-glass) phase separation at a lower density and lower temperature due to the nature of the attractive potential, which is characterized by the negative slope in the isothermal pressure-density curve~\cite{sastry2000liquid,sastry1997statistical}. This instability appears below the so-called Sastry density that locates the minimum of the pressure-density curve. Since the Sastry density increases with decreasing temperature, we have to care about the instability for our KA$_1$ and KA$_1$ models. We have confirmed that our models do not show such instability in both equilibrium and inherent structures, having the Sastry density below $\rho=1.2$. This issue could easily be avoided by simulating these models at higher densities. We choose $\rho=1.2$ to compare our results with previous studies of the original KA model. 

\section{The experimental glass transition temperature}

We define the laboratory glass transition as $\tau^{(N)}_{\alpha}(T_g^{*})/\tau_{o}$ = $10^{12}$. Since the standard MC can only provide relaxation times at most of the order of $\tau_{\alpha}/\tau_{o} \sim 10^{4}$, where $\tau_{o}(T=0.70) \sim 3 \times 10^{3}$ MC sweeps is the value of structural relaxation time at the onset of slow dynamics. Considering this computational limitation, we need to extrapolate the dynamic behaviour to locate the experimental glass transition temperature $T_g^{*}$.

We estimate the onset temperature as the deviation from the Arrhenius form and perform a detailed study in the supercooled regime using various functional fit forms to locate the range of the glass transition temperature \cite{ferry1980viscoelastic,elmatad2010corresponding,berthier2017configurational}. The first functional form is the Vogel-Fulcher-Tammann (VFT) expression~\cite{ferry1980viscoelastic},
\begin{equation}
\tau_{\alpha}(T) = \tau_{\infty} \exp \left[\frac{1}{K_{VFT} \left[ T/T_{VFT} -1 \right ]} \right] ,
\end{equation}
where $\tau_{\infty}$, $K_{VFT}$ and $T_{VFT}$ represent high-temperature relaxation time, the dynamic fragility and the finite temperature divergence, respectively. A second functional form is the parabolic fit~\cite{elmatad2010corresponding},
\begin{equation}
\tau_{\alpha}(T) = \tau_o \exp \left[ J \left( \frac{1}{T} -\frac{1}{T_{o}} \right)^{2} \right] ,
\end{equation}
where $T_{o}$ represents the onset temperature. In contrast to the VFT law, this parabolic fit form does not predict a divergence of the relaxation time at any finite temperature. The parabolic law can be seen as the simplest correction to the Arrhenius behaviour that includes dynamic fragility. 

As a result, the VFT presumably overestimates the extrapolated relaxation time, and the parabolic law is presumably a safer extrapolation. Indeed a recent study suggests that the parabolic fit form is more reliable and consistent with the experimental data when extrapolated from a small range in the computational domain~\cite{ozawa2019does}. In the main text, we mark the laboratory glass transition range from the VFT and parabolic fits obtained below $T_{o}$. These values are used as boundaries for the possible location of $T_g$.

\section{Thermodynamics}

We provide the details of the calculation of the configurational entropy, $s_{c}$, which is estimated by the computation of the total, vibrational, and mixing entropies~\cite{ozawa2018configurational}.

\subsection{Total entropy}

The total free energy $A(\rho,T)$ of the system at a density $\rho$ and temperature $T$ can be written as the sum
of the ideal gas part $A_{id}(\rho,T)$ and the excess free energy $A_{ex}(\rho,T)$.
The ideal gas part $A_{id}(\rho, T)$ can be expressed as
\begin{eqnarray}
\beta A_{id}(\rho, T) &=& N(3 \ln \Lambda + \ln \rho - 1).
\end{eqnarray}
where $\Lambda$ is the thermal wavelength. The excess free energy $A_{ex}(\rho, T)$ is estimated by thermodynamic integration from a known limit~\cite{allen2017computer,sastry2000evaluation}. We consider the ideal gas as this reference state point. The computation of the excess free energy can be performed in two steps.

{\bf Step I}: We integrate the excess pressure $P_{ex}$ from the dilute limit to the target density $\rho=1.20$ at constant temperature $T_{r}=5.00$:
\begin{equation}
\beta_{r}A_{ex}(\rho,T_{r}) = \beta_{r} A_{ex}(0,T_{r}) + N\int_0^{\rho} \frac{d\rho'}{\rho'} \left( \frac{\beta_{r} P}{\rho'} -1\right ).
\end{equation}
The excess free energy of the reference state $A_{ex}(0,T_{r})$ contains a combinatorial term resulting from the distinct particle types. This term is the same as the mixing part in the vibrational entropy and cancels out from the configurational entropy.

{\bf Step II}: The excess free energy at the desired temperature $T$, $A_{ex}(\rho,T)$, is calculated by integrating the average potential energy $U$ from temperature $T_r$ to $T$~\cite{sastry2000evaluation}:
\begin{eqnarray} \label{path2}
\beta A_{ex}(\rho,T) &=& \beta_{r} A_{ex}(\rho,T_{r}) + \int_{\beta_{r}}^{\beta} d\beta' U(\rho,\beta').
\end{eqnarray}
The total entropy $s_{tot}$ at the target state point is obtained using the relation $s_{tot} = \frac{1}{N} \frac{\delta A} {\delta T}$.

\begin{figure*}
\centering{}
\includegraphics[scale=0.44]{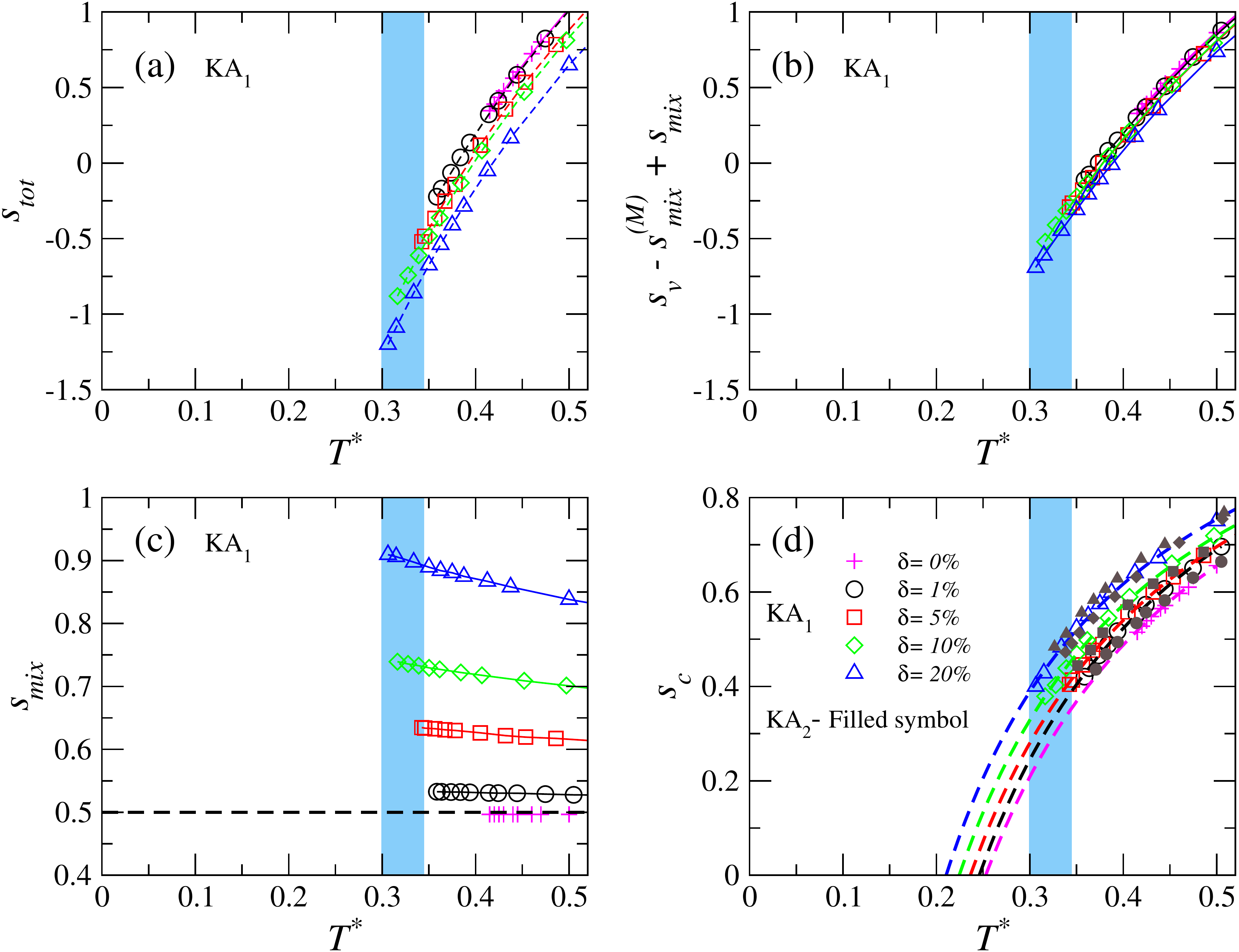}
\caption{Temperature evolution of different contributions to the configurational entropy.
(a) Total entropy $s_{tot}$.
(b) Vibrational entropy $s_{v}$.
(c) Mixing entropy. The horizontal dashed line corresponds to the combinatorial mixing entropy for the $\delta = 0\%$ KA binary mixture.
(d) Configurational entropy for the KA$_1$ and KA$_2$ models, lines are fits to the KA$_{1}$ models to estimate various thermodynamics properties.
The blue area marks the laboratory glass transition $T_{g}^{*}$.}
\label{thermodynamics}
\end{figure*}

\subsection{Vibrational entropy}

The vibrational entropy is estimated by employing the modified Frenkel-Ladd method~\cite{ozawa2018configurational}, where the Einstein solid is used as a reference state to perform a contrained thermodynamic integration. The constrained Hamiltonian is
$$
\beta U_{\alpha}(r^{N},r_{0}^{N}) = \beta U_{0}(r^{N}) + \alpha_{s} \sum_{i=1}^{N}|r_{i} -r_{i,0}|^{2},
$$
where $r_{0}^{N}$ is the reference equilibrium configuration. In the limit of large stiffness (say, $\alpha_{s,max}$) the system behaves as a classical non-interacting ensemble of harmonic oscillators. For sufficiently small values (say, $\alpha_{s,min}\rightarrow 0$) the liquid state is restored.
The resulting vibrational entropy can be expressed as~\cite{ozawa2018configurational}
\begin{eqnarray}
s_{v} &=& (3/2) - 3 \ln \Lambda - (3/2) \ln (\alpha_{s,max}/\pi) \nonumber \\
&+& \lim_{\alpha_{s,min}\rightarrow 0} \int_{\alpha_{s,min}}^{\alpha_{s,max}} d\alpha_{s}~\Delta_{\alpha_{s}}^{T,S} + s^{(M)}_{mix} - s_{mix}(r_0^{N},\beta). \nonumber \\
\label{sbasin}
\end{eqnarray}
Here, $s^{(M)}_{mix}$ is the mixing entropy stemming from the combinatorial factor of the different particle type which cancels out from the ideal gas contribution. The term $\Delta_{\alpha}^{T,S}$ is the mean-squared displacement defined by
\begin{equation}
\Delta_{\alpha_{s}}^{T,S} = \frac{1}{N} \left < \sum_{i=1}^{N} |r_{i} - r_{i,0}|^{2} \right >_{\alpha_{s}}^{T,S},
\end{equation}
where the upper scripts $T,S$ indicate both translational and swap MC displacements should be performed.
Discretisation of the integral over $\alpha$ and boundary values are chosen as in Ref.~\cite{ozawa2018configurational}.

\subsection{Mixing entropy}

The mixing entropy $s_{mix}(r_{0}^{N}, \beta)$ in Eq.~(\ref{sbasin}) needs a separate estimate~\cite{ozawa2018configurational}. We perform a thermodynamic integration over a temperature range $\beta'$ from the target temperature $\beta =1/T$ with a given reference configuration to the high temperature limit, $\beta' \rightarrow 0$. For a given configuration ($r_{0}^{N}$), the mixing entropy can be expressed as
\begin{equation}
s_{mix}(r_{0}^{N},\beta) = \frac{1}{N} \int_{o}^{\beta} d\beta' \Delta U_{mix}(r_{0}^{N}, \beta'),
\label{smixc}
\end{equation}
where $\Delta U_{mix}(r_{0}^{N}, \beta) = \left <U_0(r_{0}^{N}) \right >^{S}_{\beta'} - \left < U_{0}(r_{0}^{N}) \right >$ corresponds to the potential energy difference between the reference sample and the heated sample.
In Eq.~(\ref{smixc}), the system is heated from the initial state point $\beta$ to $\beta'\rightarrow 0$, but only particle permutations are performed while the particle positions $r_{0}^{N}$ are unchanged.

\subsection{Configurational entropy}

Collecting all terms, the configurational entropy is finally obtained as
\begin{equation}
s_{c}(T^{*})=s_{tot}(T^{*})-s_{vib}(T^{*})% +s_{mix}(T^{*}),
\end{equation}
so that $s_{vib} \equiv s_v - s_{mix}$.
We show in Fig.~\ref{thermodynamics} the temperature dependence of the various contributions to the configurational entropy as a function of $T^*$ and different values of $\delta$.

\subsection{Characteristics of KA$_2$ models} 
The table \ref{data-KA2} provides additional details for the characterization of the KA$_{2}$ models. 
\begin{table}{}
\begin{tabular}{l*{5}{c}}
KA$_{2}$              & \quad $\delta=0\%$ \quad & \quad 1\% \quad & \quad 5\% \quad & \quad 10\% \quad & \quad 20\% \\
\hline
$\varepsilon$ 					 				& $0$ & $0.01$ & $0.08$ & $0.15$ & $0.27$ \\
$T^{*}_{Low}/T^{*}_{mct}$		        & $0.954$ & $0.852$ & $0.807$ & $0.777$ & $0.750$ \\
Speedup & $1$ & $10^{2}$ & $2\times 10^{3}$ & $3\times 10^{4}$ & $2\times 10^{5}$ \\
$K_{T}$            		 					& $0.335$ & $0.329$ & $0.316$ & $0.293$ & $0.271$ \\
$T^{*}_{K}$					 				& $0.252$ & $0.249$ & $0.235$ & $0.219$ & $0.206$ \\
%$T^{*}_{Low}/T^{*}_{K}$		            & $1.647$ & $1.489$ & $1.494$ & $1.544$ & $1.584$ \\
\hline
\end{tabular}
\caption{Characteristics of the KA$_{2}$ models. Scaling factor for the temperature $\varepsilon$, the lowest simulated temperature $T^*_{low}$ relative to the mode-coupling crossover $T^*_{mct}$, the factor of thermalisation speedup, the thermodynamic fragility $K_T$, and the extrapolated Kauzmann temperatures $T^{*}_K$.} 
\label{data-KA2}
\end{table} 

\section{Adam-Gibbs relation}

\begin{figure}
\includegraphics[width=8.cm]{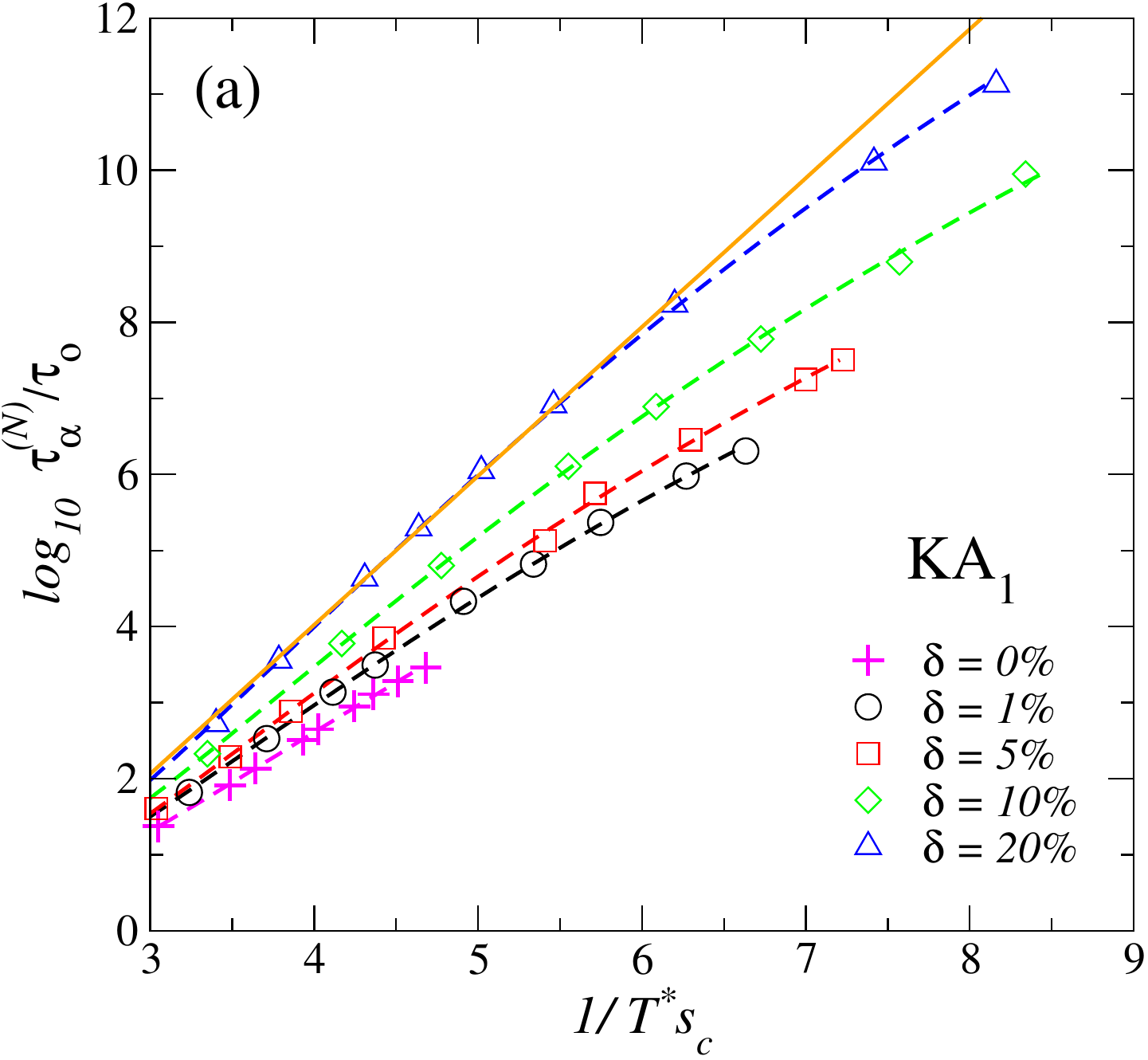}
\includegraphics[width=8.cm]{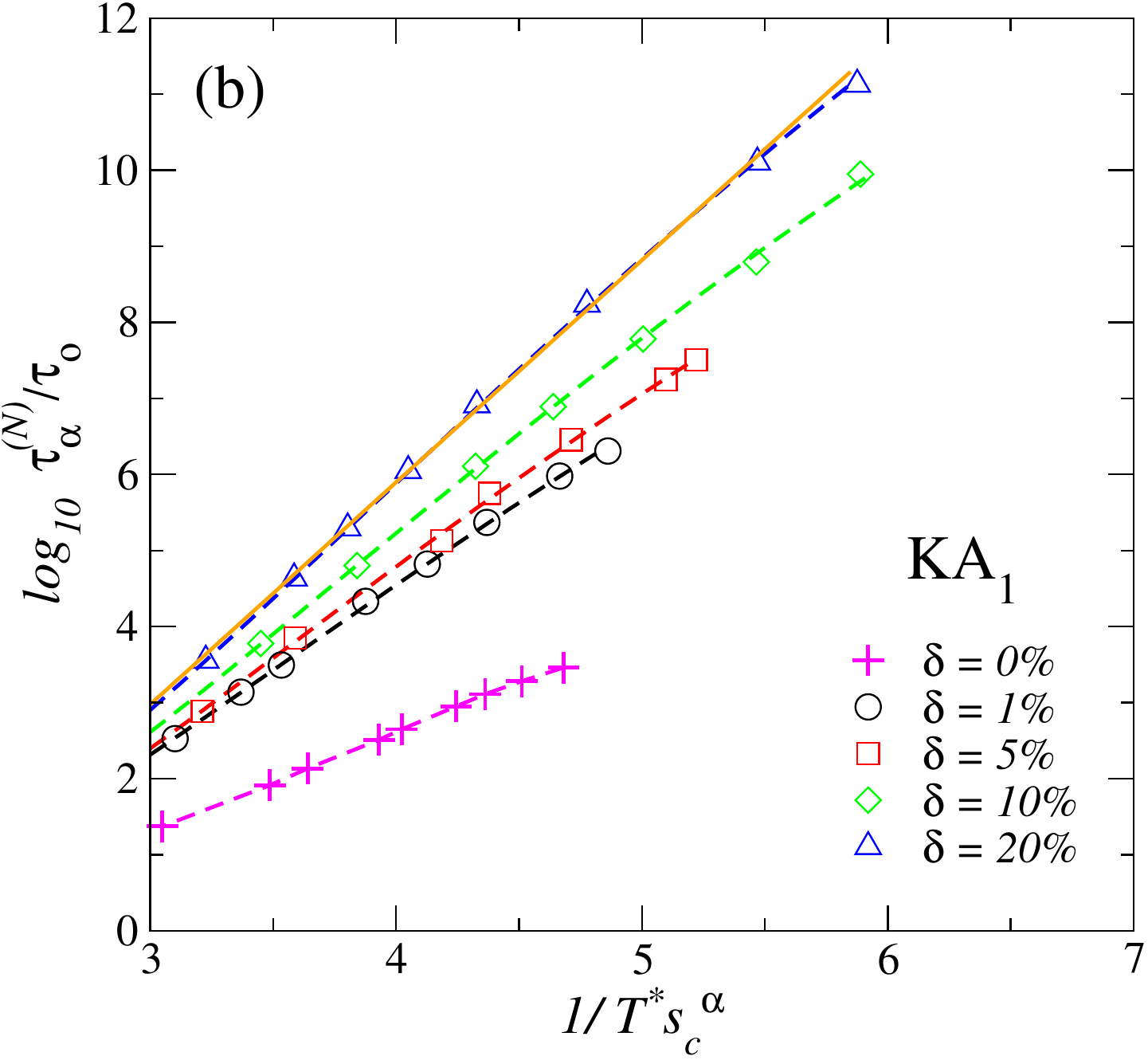}
\caption{(a) Adam-Gibbs plot for various $\delta$. The dynamics combines data measured and extrapolated using the parabolic law. The original Adam-Gibbs relation corresponds to a straight line.
(b) Generalized Adam Gibbs plot using $\alpha=0.64$ for
all $\delta$ values.}
\label{AG}
\end{figure}

We attempt to test thermodynamic theories of the glass transition
over the experimentally relevant temperature regime. The Adam-Gibbs
relation is a simple connection between thermodynamics and dynamics~\cite{adam1965temperature}. In the same vein,
the random first-order transition (RFOT) theory emphasizes the
role of configurational entropy and provides a generalised connection between thermodynamics and dynamics~\cite{kirkpatrick1989scaling,bouchaud2004adam,lubchenko2007theory}.

The RFOT theory version of the Adam-Gibbs relation can be expressed as
\begin{equation}
\log (\tau_{\alpha}/\tau_0) \propto \frac{1}{Ts_{c}^{\alpha}},
\label{eq:alpha}
\end{equation}
where $\alpha=1$ restores the original AG relation.

Recently, an extensive study of a range of simulation
and experimental measurements suggested a possible modification of
the initially proposed Adam-Gibbs relation~\cite{ozawa2019does}. We follow this recent analysis for the KA$_{1}$ models. Our results are in Fig.~\ref{AG}.

In Fig.~\ref{AG}(a), we show the relaxation times $\tau_{\alpha}^{(N)}$ combining direct measurements and extrapolations using the parabolic law, against the measured value of the configurational entropy $s_c$. The representation shows $\log_{10} \tau_\alpha$ versus $1/(T^* s_c)$ such that the original Adam-Gibbs relation would predict a straight line.

The inefficient sampling and possible crystallization of the standard KA model do not allow us to perform a detailed study of the Adam Gibbs relation at very low temperatures, and the relation seems to hold over a relatively narrow dynamic range.

Instead, the KA$_{1}$ models can be equilibrated to much lower temperatures because the swap MC algorithm is more efficient for them. The data in Fig.~\ref{AG}(a) suggest that when considered over a broader dynamic range, systematic deviations from the Adam Gibbs relation become clearly visible.

As found before for other models~\cite{ozawa2019does}, we find these small deviations from the Adam Gibbs relation can be accounted for using an exponent $\alpha < 1$ in Eq.~(\ref{eq:alpha}). In Fig.~\ref{AG}(b), we show that an exponent $\alpha \approx 0.64$ actually describes the behaviour of all models from $\delta=0\%$ (the original KA model) up to $\delta=20\%$. Values of $\alpha$ systematically smaller than 1 were also reported in many materials~\cite{ozawa2019does}.

\section{Glass forming ability}

\begin{figure}
\includegraphics[width=8.5cm]{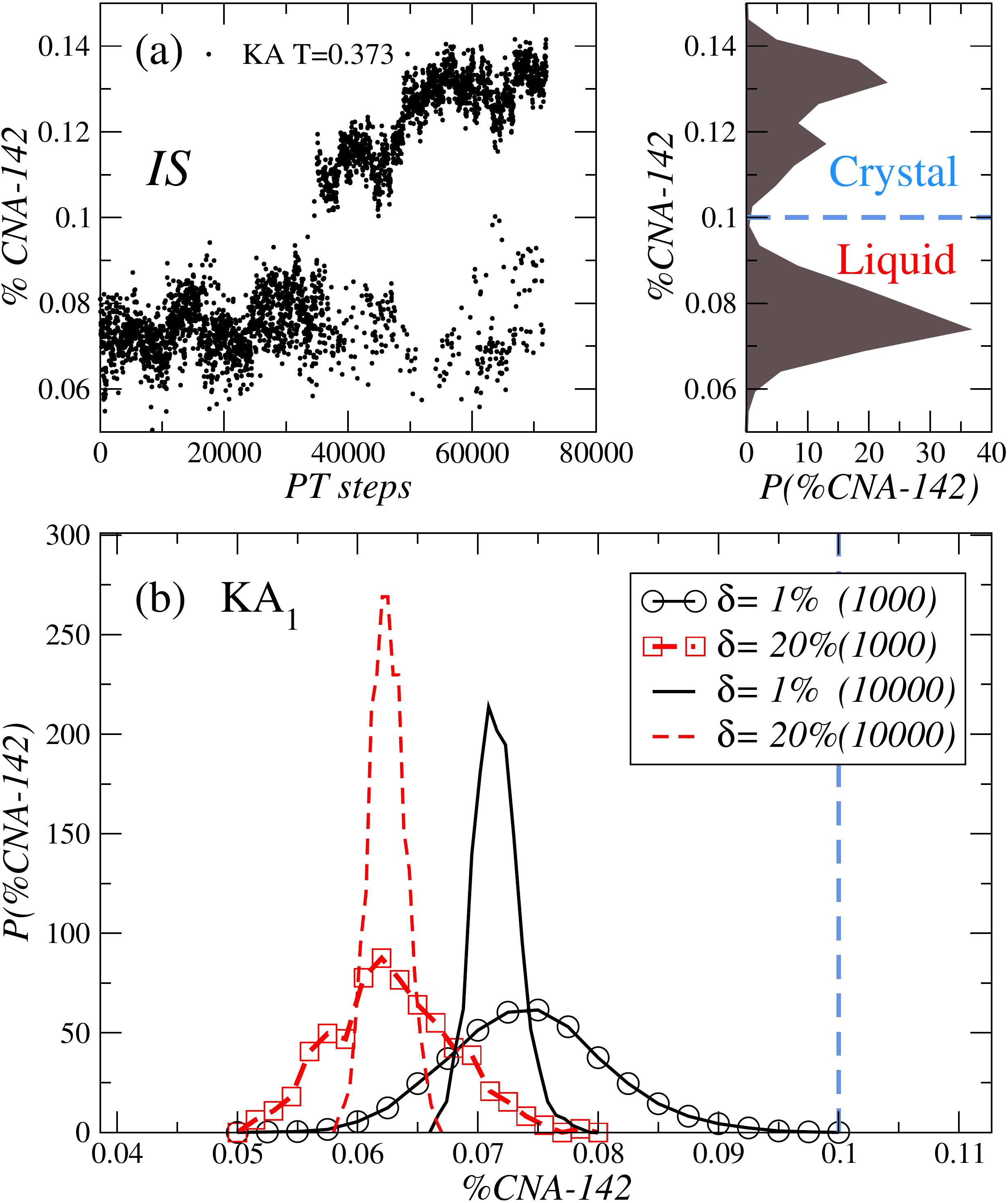}
\caption{The fraction of CNA-142 bonds, for the inherent structures, as a function of PT MD steps at $T=0.373$ and $N=1200$ from Ref.~\cite{coslovich2018dynamic}. The distribution is bimodal and suggests that 10\% is the boundary between crystal ($>$10\%) and liquid.
(b) The same distribution for $\delta=1\%$ and 20\% for $T^{*}=0.359$ and $T^{*}=0.306$, respectively. For each $\delta$, two system sizes, $N=1000$ and $10000$ are shown. The KA$_1$ model shows no sign of crystallisation.}
\label{CNA}
\end{figure}

Although considered a good glass-former for many years, progress in computer resources has led to the conclusion that the model
can be relatively easily crystallised at low enough temperatures~\cite{toxvaerd2009stability,coslovich2018dynamic,ingebrigtsen2019crystallization}. The KA mixture first demixes, and the large A particles partially crystallize into a face-centered cubic (FCC) structure~\cite{coslovich2018dynamic,ingebrigtsen2019crystallization}.

As an indicator of crystallization events in the simulations, we focus on the common neighbor analysis (CNA). In this strategy, the bonds formed by neighboring particles are quantified according to the number of shared neighbors, and therefore, the local crystalline structure provides a specific signature~\cite{honeycutt1987molecular}. We perform CNA analysis for the inherent structures of the KA model generated at $T=0.373$ using parallel tempering (PT) simulation scheme (trajectories taken from Ref.~\cite {coslovich2018dynamic}). To define neighbors, we consider a cutoff of $r_{cut}^{AA}=1.4\sigma_{AA}$, which is the minimum of the first coordination shell of the pair correlation $g_{AA}(r)$. Since the majority of the population is of type A, this is a reasonable choice. At low temperatures, the standard KA model is known to form local FCC crystals, which is well characterized as ``$142$" in the CNA analysis.

In Fig.~\ref{CNA}(a), we report the fraction of FCC population as a function of PT steps and their distribution for $T=0.373$, which provides an upper bound for the fraction \%CNA-142 in the liquid near $0.10$.

In Fig.~\ref{CNA}(b), we report the probability distribution of the fraction of FCC particles at the lowest simulated temperatures for $\delta=1\%$ and 20\% in the KA$_{1}$ model. The distribution remains well below the 0.10 cutoff, suggesting that 1\% of impurity is enough to make the system very robust against crystallization. Moreover, we confirm that a larger system size, $N=10000$, does not show the trend of crystallization. To strengthen this point, we emphasize that 30 independent swap MC runs have been performed for the 1\% model for a duration of about $15~\tau^{(S)}_{\alpha}$ at the lowest temperature ($T=0.355$) and $N=1000$ particles. 
For $N=10000$ particle case for the 1\% model, we consider 10 independent swap MC runs of duration $\approx1.5~\tau^{(S)}_{\alpha}$ at the lowest temperature ($T=0.355$).
None of them show any signatures of crystallization or demixing.

%@@ What about $N=10000$ particles data? @@ 

%\bibliography{metallic-glasses.bib}

\end{document}